\documentstyle[12pt]{article}
\def\cw{\cos\theta_W}
\def\gev{\; {\rm GeV}}
\def\gh{\Gamma ( H \rightarrow h h )}

\def\mt{m_t}
\def\mb{m_b}
\def\ma{m_A}
\def\matb{(m_A,\tan \beta)}

\def\msta{m_{\tilde{t}_1}}
\def\mstb{m_{\tilde{t}_2}}
\def\mw{m_W}
\def\mz{m_Z}
\def\mhh{m_H}
\def\mh{m_h}
\def\mhc{m_{H^{\pm}}}
\def\ov{\overline}

\def\sqa{\tilde{q}_a}
\def\sqb{\tilde{q}_b}
\def\sqc{\tilde{q}_c}
\def\simlt{\mathrel{\lower2.5pt\vbox{\lineskip=0pt\baselineskip=0pt
           \hbox{$<$}\hbox{$\sim$}}}}
\def\simgt{\mathrel{\lower2.5pt\vbox{\lineskip=0pt\baselineskip=0pt
           \hbox{$>$}\hbox{$\sim$}}}}
\def\s2w{\sin 2 \theta_W}
\def\tb{\tan \beta}
\def\tev{\; {\rm TeV}}
\def\to{\rightarrow}

\def\v1t{V^{(1)}[m^2(\phi),T]}
\newcommand{\be}{\begin{equation}}
\newcommand{\ee}{\end{equation}}
\newcommand{\bea}{\begin{eqnarray}}
\newcommand{\eea}{\end{eqnarray}}
\newcommand{\bean}{\begin{eqnarray*}}
\newcommand{\eean}{\end{eqnarray*}}
\begin{document}
\begin{titlepage}
\vspace*{-1cm}
\noindent
\phantom{DRAFT 15/10/92}
\hfill{CERN-TH.6603/92}
\newline
\phantom{DRAFT 15/10/92}
\hfill{hep-ph/9210266}
\vskip 2.5cm
\begin{center}
{\Large\bf Radiative corrections to the decay $H \to h h$ in}
\\
\vskip 0.5cm
{\Large\bf the Minimal Supersymmetric Standard Model}
\end{center}
\vskip 1.5cm
\begin{center}
{\large A. Brignole}\footnote{Also at Dip. di Fisica,
Univ. di Padova, and INFN, Sezione di Padova, Padua, Italy.}
\\
\vskip .3cm
and \\
\vskip .3cm
{\large F. Zwirner}\footnote{On leave from INFN, Sezione di Padova, Padua,
Italy.}
\\
\vskip .3cm
Theory Division, CERN, \\
Geneva, Switzerland \\
\vskip 1cm
\end{center}
\begin{abstract}
\noindent
We set up a suitable renormalization programme for the one-loop computation
of the decay rate $\gh$ in the Minimal Supersymmetric extension of the
Standard Model.
We then perform an explicit diagrammatic calculation, including the full
contributions from top, bottom, stop and sbottom loops.
We find that, for $\tb$ close to 1 and $\mhh \simgt 2 m_t$, the results
can significantly differ from those previously obtained in the effective
potential approach.
However, the latter method remains a good approximation in the region of
parameter space which is most relevant for $H$ searches at large hadron
colliders.
\end{abstract}
\vfill{
CERN-TH.6603/92
\newline
\noindent
July 1992}
\end{titlepage}
\vskip2truecm

\setcounter{footnote}{0}

Radiative corrections to the parameters of the Higgs boson
sector in the Minimal Supersymmetric extension of the Standard
Model (MSSM) have recently received much attention.

After the discovery [\ref{pioneer}] that top and stop loops can
cause large corrections to the masses of the neutral CP-even
Higgs bosons\footnote{Previous studies [\ref{previous}] either
neglected the case of a heavy top quark, or concentrated on the
violations of the neutral Higgs mass sum rule, without computing
corrections to the individual Higgs masses.}, radiative corrections
to Higgs boson masses have been computed by a variety of methods: the
renormalization group approach [\ref{rga}], the effective potential
approach [\ref{epa}--\ref{drees}] and the diagrammatic approach
[\ref{da}--\ref{b2}].

The renormalization group approach assumes that there are two (or more)
widely separated mass scales, for example
\be
\label{brute}
M_{SUSY} \; (\sim \msta \sim \mstb \sim \ldots \sim \mhh \sim \mhc
\sim \ma) \gg \mz \; (\sim \mh \sim \mt) \, ,
\ee
and considers the effective theory for the degrees of freedom lighter
than $M_{SUSY}$. It then solves (non-supersymmetric) renormalization
group equations to obtain running parameters down to the scale $Q = \mz$,
imposing the tree-level relations of the MSSM as boundary conditions at
the scale $Q = M_{SUSY}$. This approach has the
advantage of resumming the leading corrections, proportional to
$\log(M_{SUSY}/\mz)$, so that even the case of $M_{SUSY}$ orders of
magnitude larger than $\mz$ can be dealt with in perturbation theory.
On the other hand, if supersymmetry is to solve the naturalness problem
of the Standard Model, one expects the various mass parameters of the MSSM
to be scattered around the electroweak scale, $G_F^{-1/2} \simeq 250 \gev$,
so that assumption (\ref{brute}) breaks down.

The effective potential approach consists in identifying the Higgs boson
masses and self-couplings with the corresponding derivatives of the
one-loop effective potential, evaluated at the minimum. By definition,
this approach evaluates all Higgs self-energies and vertices at vanishing
external momentum. In the case of radiative corrections to the Higgs boson
masses, this was shown to be a rather accurate approximation
[\ref{b1},\ref{b2}].
Actually, when the external momentum (i.e. the Higgs mass) approaches
or exceeds the threshold of the internal particles, the full
correction can be rather different from the zero-momentum one. However,
in that case corrections themselves are small, either in the absolute sense
or relatively to the (increased) tree-level mass.
Other possible drawbacks of the effective potential approach are the
gauge- and scale- dependence of the associated quantities\footnote{We
recall that, although the effective potential is scale-independent,
scale dependence enters its derivatives through the renormalized fields.
More generally, the issue involved here is the dependence on the
renormalization scheme, scale dependence being interpretable as a
particular kind of scheme dependence.}. These are not
serious problems in the computation of the mass corrections:
the dominant ones come from quark and squark loops, which do not
introduce spurious dependences on the gauge parameter into the results;
also, wave-function renormalization effects, responsible for the scale
dependence, are generally small with respect to the overall mass corrections.

The diagrammatic approach consists in performing the complete
one-loop renormalization programme, specifying unambiguously the input
parameters and the relations between renormalized parameters and physical
quantities. This approach gives the most precise computational
tool in the case of supersymmetric particle masses spread around the
electroweak scale, and results which are formally gauge- and
scale-independent. Since corrections can be numerically large, however,
one has to pay attention and improve conveniently the na\"{\i}ve one-loop
calculations when necessary. An example is the determination of the
neutral CP-even masses, as discussed in detail in ref.~[\ref{b2}].

Whilst radiative corrections to Higgs boson masses are by now well
under control, the study of radiative corrections to Higgs boson
couplings is still at a less refined stage. In most
phenomenological [\ref{epa},\ref{berz},\ref{kz}--\ref{pheno}] and
experimental [\ref{exper}] studies, radiative corrections to the Higgs
couplings to vector bosons and fermions have been taken into account
only approximately, by improving the tree-level formulae with one-loop
corrected values of the $h$--$H$ mixing angle, $\alpha$, and with running
fermion masses, evaluated at the typical scale $Q$ of the process under
consideration. Residual corrections are expected to be numerically small,
with the possible exception of important threshold effects
[\ref{polish},\ref{pp}].

In the case of the Higgs boson self-couplings, which control decays like
$H \to hh$, $H \to AA$ and $h \to AA$ when the latter are kinematically
allowed, it is known [\ref{berz}] that radiative corrections can be
numerically large. Radiative corrections to cubic Higgs boson self-couplings
have been computed, at different levels of approximation, both in the
effective potential approach [\ref{berz},\ref{b3},\ref{bbps}] and
in the renormalization group approach [\ref{hhn}]. Given the fact that,
in addition to the masses of the virtual particles in the one-loop diagrams,
two different mass scales are involved in the decays $H \to hh$, $H \to AA$
and $h \to AA$, the mass of the decaying particle and the mass of the decay
products, one might suspect
that momentum-dependent effects, which are neglected in the renormalization
group and in the effective potential approaches, could play a role.

The purpose of the present work is to perform the diagrammatic
calculation of $\Gamma(H \to hh)$ at the one-loop level, and to compare
the results with those obtained in other approaches. The main motivations
for choosing this particular decay are the relative simplicity of the
calculation and the fact that, even after the inclusion of the leading
radiative corrections, $H \to h h $ is the dominant $H$ decay mode over
a large region of parameter space. A detailed discussion of the MSSM Higgs
branching ratios at the one-loop level would require extending the present
calculation to other decay modes, as currently under study.

This paper is organized as follows. We begin by setting up a convenient
one-loop renormalization programme for the decay rate $\gh$ in the MSSM.
We then perform a complete computation of the contributions due to top,
stop, bottom and sbottom loops. Finally, we compare our results with those
obtained at the tree level and in the effective potential approach, and we
discuss their phenomenological implications for the detection of $H$ at
future colliders.


The notation of the present paper will closely follow that of
[\ref{b1},\ref{b2}], unless otherwise stated. In [\ref{b2}],
radiative corrections to the neutral Higgs boson masses were
computed in the $\ov{DR}$ scheme [\ref{drbar}], using the physical
mass $\ma$ and $\beta \equiv \beta^{\ov{DR}} (Q=\mz)$ as input
parameters, and explicit formulae for the physical masses $\mh$
and $\mhh$ were given. Here we shall adopt the same
renormalization scheme for the computation of  $\Gamma(H \to hh)$
at the one-loop level. We define the $H$ and $h$ fields in the CP-even
neutral Higgs sector by $H =  \cos\alpha \; S_1 + \sin\alpha \; S_2$ and
$h = -\sin\alpha \; S_1 + \cos\alpha \; S_2$. As explained in [\ref{b2}],
for a satisfactory convergence of the perturbative expansion it is important
to define the mixing angle $\alpha$ in terms of a mass matrix which
includes the leading, momentum-independent one-loop self-energy corrections.

The decay rate for the process under consideration reads
\be
\label{rate}
\Gamma(H \to hh) = \frac{|{\cal A}(H \to hh)|^2}{32\pi\mhh}
\sqrt{1-\frac{4\mh^2}{\mhh^2}} \;\; ,
\ee
where the one-loop-corrected  amplitude $\cal A$ is given by
\be
\label{ampl}
{\cal A} = \sum_{i,j,k} (Z_H^{1/2})_{Hi} \, (Z_h^{1/2})_{hj} \,
 (Z_h^{1/2})_{hk} \, \lambda_{ijk} + \Lambda_{Hhh}  \;\; .
\ee
In eq. (\ref{ampl}), $i,j,k=h,H$ and $\lambda_{ijk}$ are the tree-level
cubic Higgs couplings in the CP-even sector\footnote{In the Feynman rules
these couplings are multiplied by $i$. Also, we denote by $i\Lambda$
the proper vertex obtained from the Feynman diagrams.}
\bea
\lambda_{HHH} & = & -\frac{g\mz}{2\cw} 3\cos(\beta+\alpha)\cos(2\alpha) \; ,
\\
\lambda_{HHh} & = & -\frac{g\mz}{2\cw}
[-2 \cos(\beta+\alpha)\sin(2\alpha) - \sin(\beta+\alpha)\cos(2\alpha)]  \; ,
\\
\lambda_{Hhh} & = & -\frac{g\mz}{2\cw}
[2 \sin(\beta+\alpha)\sin(2\alpha) - \cos(\beta+\alpha)\cos(2\alpha)]  \; ,
\\
\lambda_{hhh} & = & -\frac{g\mz}{2\cw}
3\sin(\beta+\alpha)\cos(2\alpha) \; ,
\eea
where $g$, $\mz \equiv \sqrt{(g^2 + g'^2) (v_1^2 + v_2^2)/2}$ and $\cw
\equiv g /\sqrt{g^2 + g'^2}$ are ($\ov{DR}$) renormalized parameters.
The last term $\Lambda_{Hhh}$ is the  ($\ov{DR}$) renormalized one-loop
proper vertex, evaluated with on-shell external momenta.

Finally, the $2\times 2$ matrices $Z_H$ and  $Z_h$ correspond to
a finite wave function renormalization, which must be taken into
account in the computation of a physical amplitude.
Such corrective factors can
be interpreted as the effect of inserting `bubbles'
on the external legs of the tree-diagrams. More precisely, the
two matrices $Z_H$ and $Z_h$ are nothing else than the (matrix-)
residues of the renormalized (matrix-) propagator $G_{ij}(p^2)$
of the neutral CP-even Higgses, at the poles $p^2=m_H^2$ and
 $p^2=m_h^2$, respectively. Actually, $Z_H$ and $Z_h$ are
extracted from the real part of the propagator, so in the
following we shall implicitly consider only real parts of
propagators and self-energies.
The renormalized inverse propagator
$\Gamma(p^2)$ was computed, in the $(S_1,S_2)$ basis,
in ref. [\ref{b2}], and the one-loop-corrected masses $\mhh$
and $\mh$ were extracted from it. We now rewrite  $\Gamma(p^2)$
in the $(H,h)$ basis in terms of $\mhh$ and $\mh$:
\be
\label{gammaij}
\Gamma_{ij}(p^2) =
\left( \begin{array}{cc}
p^2 - \mhh^2 +\hat\Pi_{HH}(p^2)-\hat\Pi_{HH}(\mhh^2) & \Gamma_{Hh}(p^2)
\\
\Gamma_{hH}(p^2) & p^2 - \mh^2 +\hat\Pi_{hh}(p^2)-\hat\Pi_{hh}(\mh^2)
\end{array}  \right)
\, .
\ee
In eq.~(\ref{gammaij}), the $\hat\Pi_{ij}(p^2)$ are one-loop
$\ov{DR}$-renormalized self-energies, and
\be
\label{offd}
\Gamma_{Hh}(p^2)= \Gamma_{hH}(p^2)=\frac{1}{2} \sin2(\beta+\alpha)
\hat\Pi_{ZZ}(\mz^2) +\frac{1}{2} \sin2(\beta-\alpha)
\Delta\hat\Pi_{AA}(\ma^2)+\Delta\hat\Pi_{Hh}(p^2) \; ,
\ee
where for a generic self-energy $\hat\Pi(p^2)$ we define $\Delta
\hat\Pi(p^2) \equiv \hat\Pi(p^2) - \hat\Pi(0)$. Around the pole
$p^2=\mhh^2$, we have $G(p^2)\sim Z_H/(p^2-m_H^2)$, where
\be
\label{zhh}
Z_H =
\left( \begin{array}{cc}
1-\hat\Pi'_{HH}(\mhh^2) & -\Gamma_{Hh}(\mhh^2)/(\mhh^2-\mh^2)
\\
-\Gamma_{Hh}(\mhh^2)/(\mhh^2-\mh^2) & 0
\end{array} \right)
\, .
\ee
Around the pole $p^2=\mh^2$, we have $G(p^2)\sim Z_h/(p^2-\mh^2)$, where
\be
\label{zh}
Z_h =
\left( \begin{array}{cc}
0 & -\Gamma_{Hh}(\mh^2)/(\mh^2-\mhh^2)
\\
-\Gamma_{Hh}(\mh^2)/(\mh^2-\mhh^2) & 1-\hat\Pi'_{hh}(\mh^2)
\end{array} \right)
\, .
\ee
In eqs.~(\ref{zhh}) and (\ref{zh}), $\hat\Pi' (p^2) \equiv  d \hat\Pi (p^2)
/ d p^2$.

Expanding eq. (\ref{ampl}), we finally obtain
\be
\label{aform}
{\cal A}  =
\left[ 1-\frac{1}{2}\hat\Pi'_{HH}(\mhh^2)-\hat\Pi'_{hh}(\mh^2) \right]
\lambda_{Hhh}
+ \frac{2\Gamma_{Hh}(\mh^2)\lambda_{HHh}-\Gamma_{Hh}(\mhh^2)\lambda_{hhh}}
{\mhh^2-\mh^2} + \Lambda_{Hhh} \,  .
\ee


We now proceed to the evaluation of the general formula for $\cal A$,
eq. (\ref{aform}), in the particular case in which only diagrams
corresponding to top-stop-bottom-sbottom loops are taken into account.
This should give the dominant one-loop correction to $\Gamma(H\to hh)$
[\ref{hhn}].

The expressions for $\hat\Pi'_{HH}(\mhh^2)$ and $\hat\Pi'_{hh}(\mh^2)$
can be trivially obtained from the expressions of $\Delta\hat\Pi_{HH}$
and $\Delta\hat\Pi_{hh}$ given in ref. [\ref{b2}].
To obtain an explicit expression for $\Gamma_{Hh}(p^2)$, eq. (\ref{offd}),
we need $\hat\Pi_{ZZ}(\mz^2)$, $\Delta\hat\Pi_{AA}(\ma^2)$ and
$\Delta\hat\Pi_{Hh}(p^2)$. The first two were given\footnote{The expression
of $\hat\Pi_{ZZ}(\mz^2)$, which was correctly given in the preprint version,
contains a misprint in the published version: the coefficient appearing in
the third line should read $(s^2_t c_{tL} - c^2_t c_{tR})^2$ and not
$(s^2_t c_{tL} + c^2_t c_{tR})^2$. There are other misprints in the published
version, which were not present in the preprint version: in the conventions
stated in footnote number 2, the part of a gauge-boson self-energy diagram
proportional to $g_{\mu\nu}$ should read $-i g_{\mu\nu} \Pi(p^2)$; in the
expression for $\Delta'_{11}/m_Z^2$, $d^t_{12}$ should read $2 d^t_{12}$;
the remaining ones are obvious.} in ref. [\ref{b2}]. For the last one we find:
\bea
\label{delta}
\Delta\hat\Pi_{Hh}(p^2) & = &
\frac{3g^2\mt^2}{16\pi^2\mw^2} \frac{\sin\alpha\cos\alpha}{\sin^2\beta}
\left[ \frac{p^2}{6}+3\mt^2\Delta F(\mt,\mt,p) -3 G(\mt,\mt,p) \right]
\nonumber\\
& & -
\frac{3g^2\mb^2}{16\pi^2\mw^2} \frac{\sin\alpha\cos\alpha}{\cos^2\beta}
\left[ \frac{p^2}{6}+3\mb^2\Delta F(\mb,\mb,p) -3 G(\mb,\mb,p) \right]
\nonumber\\
& & -
\frac{3g^2}{16\pi^2\mw^2} \sum_{q=t,b}\sum_{a,b=1,2}
c_{H\sqa\sqb}c_{h\sqa\sqb} \Delta F(m_{\sqa},m_{\sqb},p)  \;\; .
\eea
In eq.~(\ref{delta}), $p \equiv \sqrt{p^2}$ and
$\Delta F (m_1, m_2, p) \equiv F (m_1, m_2, p)
- F (m_1, m_2, 0)$, whilst the coefficients $c_{i\sqa\sqb}$ correspond
to the trilinear Higgs-squark-squark couplings and are summarized in
the Appendix. The functions $F$ and $G$ were given in [\ref{b1}].

Finally, we need to compute $\Lambda_{Hhh}$. The three basic topologies
of the diagrams contributing to $\Lambda_{Hhh}$ are depicted in fig.~1.
Accordingly, the result for $\Lambda_{Hhh}$ can be written as the sum
of three contributions:
\be
\Lambda_{Hhh}=\Lambda_{Hhh}^{(I)}+\Lambda_{Hhh}^{(II)}
+\Lambda_{Hhh}^{(III)} \, ,
\ee
where
\bea
\Lambda_{Hhh}^{(I)} & = &
\frac{3g^3\mt^4}{16\pi^2\mw^3} \frac{\sin\alpha\cos^2\alpha}{\sin^3\beta}
\left[ F^c(\mt,\mt,\mhh)+ 2 F^c(\mt,\mt,\mh) \phantom{\frac{1}{2}} \right.
\nonumber\\
& & + \left. \left( 4\mt^2 - \frac{1}{2}\mhh^2-\mh^2 \right)
f(\mt,\mt,\mt;\mhh,\mh,\mh) \right]
\nonumber\\
& & + \frac{3g^3\mb^4}{16\pi^2\mw^3} \frac{\cos\alpha\sin^2\alpha}{\cos^3\beta}
\left[ F^c(\mb,\mb,\mhh)+ 2 F^c(\mb,\mb,\mh) \phantom{\frac{1}{2}} \right.
\nonumber\\
& & + \left. \left( 4\mb^2 - \frac{1}{2}\mhh^2-\mh^2 \right)
f(\mb,\mb,\mb;\mhh,\mh,\mh) \right] \, ,
\\
\Lambda_{Hhh}^{(II)} & = &
\frac{2 \!\cdot\! 3 g^3}{16\pi^2\mw^3} \sum_{q=t,b} \;\; \sum_{a,b,c}
c_{H\sqa\sqb}c_{h\sqa\sqc}c_{h\sqb\sqc}
f(m_{\sqa},m_{\sqb},m_{\sqc};\mhh,\mh,\mh) \, ,
\\
\Lambda_{Hhh}^{(III)} & = &
\frac{-3 g^3}{16\pi^2\mw^3} \sum_{q=t,b} \;\;  \sum_{a,b}
\left[  c_{H\sqa\sqb}c_{hh\sqa\sqb}F^c(m_{\sqa},m_{\sqb},\mhh)
\right. \nonumber \\
& & + \left.
2c_{h\sqa\sqb}c_{Hh\sqa\sqb}F^c(m_{\sqa},m_{\sqb},\mh) \right] \, .
\eea
The expressions for the functions $F^c$ and $f$, as well as for
the coefficients $c_{ij\sqa\sqb}$ corresponding to the quartic
Higgs-Higgs-squark-squark couplings, are collected in the Appendix.


This concludes our analytical evaluation of the one-loop-corrected
amplitude and decay width, eqs.~(\ref{rate}) and (\ref{ampl}).
As a check equivalent to divergence cancellation, we have explicitly
verified that the amplitude, and consequently the width, do not depend on the
renormalization scale $Q$, as expected for physical quantities.
Actually, such $Q$ independence holds up to higher-order terms,
consistently with the one-loop accuracy of our computation.
It is the result of a cancellation between the explicit $Q$ dependence
of self-energies and proper vertices, and the implicit $Q$ dependence of
the parameters contained in $\lambda_{Hhh}$: in our case, consistency
requires that we consider only the $Q$ dependence of parameters that is
due to top, stop, bottom and sbottom virtual effects. In the following
numerical calculations, we set $Q=m_Z$.

Consistency would also require that we specify the input parameters $g$,
$m_Z$ and $\cw$ [or, equivalently, $\alpha \equiv g^2 \sin^2 \theta_W /
(4 \pi)$, $m_Z$ and $\cw$] in the $\ov{DR}$ scheme, and in a theory
containing the stop and sbottom degrees of freedom besides the Standard
Model particles. The $\ov{DR}$ mass $m_Z$ is related to the physical mass
$m_{Z, phys}$ by
\be
m_Z^2 = m_{Z,phys}^2 + \hat\Pi_{ZZ}(\mz^2) \, .
\ee
Similarly, the $\ov{DR}$ fine structure constant $\alpha$ is related to
$\alpha_{em}^{-1} \simeq 137$ by
\be
\frac{1}{\alpha} = \frac{1}{\alpha_{em}}
- \frac{\Delta \alpha_{light}}{\alpha}
+ \frac{1}{2 \pi} \sum_{i \in heavy} b_i \log \frac{m_i}{m_Z}
\, ,
\ee
where $\Delta \alpha_{light}$ is the (large) contribution from charged
leptons and the five observed quarks, and the remaining (small) contributions
from the heavy particles in the model, denoted by the index $i$, are
proportional to the corresponding one-loop QED $\beta$-function coefficient
$b_i$. Finally, the $\ov{DR}$ electroweak mixing angle is given
by
\be
\cos^2 \theta_W = {m_{W,phys}^2 \over m_{Z,phys}^2}
\left[ 1 +  {\hat\Pi_{WW}(\mw^2) \over \mw^2}
- {\hat\Pi_{ZZ}(\mz^2) \over \mz^2} \right] \, .
\ee
At the level of accuracy of our numerical examples, however, it is enough
to work with the fixed input parameters $m_Z = 91 \gev$, $\sin^2 \theta_W
= 0.23$ and $\alpha^{-1} = 128$. This approximation is justified by the fact
that there are other effects not accounted for in our results: one-loop
corrections involving loops of gauge bosons, Higgs bosons, gauginos and
higgsinos; also, order $h_t^2$ or $g_s^2$ (two-loop) corrections to the
one-loop diagrams considered here.

Before moving to the numerical evaluation of our results, we would like to
relate them to the results one obtains in the effective potential
approach, which consists in approximating the amplitude ${\cal A}$ by
\be
\label{aepp}
{\cal A}^{e.p.} \equiv - \left( {\partial^3 V_{eff} \over
\partial H \partial h \partial h} \right)_{min} \, .
\ee
Since this amounts to computing the $Hhh$ 3-point function at vanishing
external momenta, we can easily obtain an explicit expression for
${\cal A}^{e.p.}$ by taking a special limit of our previous result:
\be
\label{aep}
{\cal A}^{e.p.} = \lambda_{Hhh} + \left. \Lambda_{Hhh}
\right|_{m_h=0,m_H=0} \, .
\ee
As a check, we have computed ${\cal A}^{e.p.}$ from its definition
(\ref{aepp}) and verified that eq.~(\ref{aep}) gives an identical
result.

To illustrate our results, we show in fig.~2 the one-loop-corrected width
$\Gamma (H \to hh)$, as a function of $\mhh$, corresponding to four
representative parameter choices. For simplicity, in our numerical examples
we take as soft supersymmetry-breaking parameters $\tilde{m}_Q = \tilde{m}_T
= \tilde{m}_B \equiv m_{sq}$ and $A_t = A_b \equiv A$, in the conventions of
refs.~[\ref{b1},\ref{b2}]. For comparison, we also show the values of the
width obtained by replacing\footnote{In the evaluation of $\Gamma (H \to hh)$,
one also needs the masses $m_h$ and $m_H$ as functions of the input
parameters: for those we use the one-loop-corrected expressions in all
three cases. To be consistent with this prescription, even when evaluating
${\cal A}^{tree}$ we use the one-loop-corrected expression of the mixing
angle $\alpha$.} the amplitude ${\cal A}$ in eq.~(\ref{ampl}) with its
`improved tree-level' expression, ${\cal A}^{tree} = \lambda_{Hhh}$, and with
the effective-potential expression, ${\cal A}^{e.p.}$ in eq.~(\ref{aep}).
The behaviour of the one-loop-corrected width $\gh$ in the $\matb$ plane
is illustrated in fig.~3, which displays contours of constant width,
for the representative parameter choice $\mt = 140 \gev$,
$m_{sq} = 1 \tev$, $A = \mu = 0$.

As a first comment, we observe that there is a region of the $\matb$ plane
in which the decay $H \to hh$ is kinematically forbidden. At tree level,
this region corresponds to $|\cos 2 \beta| \geq 2(m_A^2+m_Z^2)/(5 m_A m_Z)$,
which implies $m_Z/2 \leq m_A \leq 2 m_Z$ and $\tan\beta \geq 3$. As
expected, this region is deformed by the inclusion of radiative corrections,
in a way which depends on $\mt$,$m_{sq}$,\ldots. It is delimited
by the thick solid line in fig.~3, and its existence is also evident
in figs.~2b and 2c. For a given $\tb$ in the above range, the forbidden
region for $H \to hh$ essentially corresponds to $m_H^{min} < m_H
\simlt 2 m_h^{max}$, where $m_H^{min}$ ($m_h^{max}$) is the lowest (highest)
possible value of $m_H$ ($m_h$).  We also recall that
the small region of $m_H \sim m_H^{min}$, corresponding to $m_A \simlt
50 \gev$, is almost entirely ruled out by the present LEP data [\ref{exper}].
{}From figs.~2a, 2d and 3 we can also see that there is an additional
line in the $\matb$ plane where $\gh$ vanishes, due to the vanishing of
the amplitude. For the parameter choice of fig.~3, this occurs for $\tb
\simlt 2$ and $\ma \sim m_W$, corresponding to $m_H \sim m_H^{min}$.

The general behaviour of $\gh$ in the $\matb$ plane is well represented
in fig.~3. For $\ma \simlt 2 m_Z$, and in the kinematically allowed
region, the decay rate depends mildly on $\tb$, and rapidly decreases
with $\ma$ approaching the critical line near $m_W$. For $m_A \simgt
2 m_Z$, the partial width has a milder dependence on $\ma$ and a stronger
dependence on $\tb$: the largest values are obtained for $\tb \sim
2{\rm -}3$ and $\ma \sim 200{\rm -}350 \gev$. For very large values
of $\tb$ the width becomes negligibly small in comparison with the competing
channnels, in particular $H \to b \ov{b}$.

As for the dependence of the corrections on $m_t$, $A$ and $\mu$, in
general $\Gamma(H \to hh)$ rapidly increases with increasing $m_t$,
$A$ and $\mu$. For example, for $m_A = 500 \gev$, $\tb = 1.5$ and
$m_{sq} = 750 \gev$, one obtains
$$
\begin{array}{lcl}
m_t = 120 \gev \, , A=\mu = 250 \gev \, ,
&
\Rightarrow
&
\Gamma = 0.04 \gev \, ;
\\
m_t = 120 \gev \, , A=\mu = 1 \tev \, ,
&
\Rightarrow
&
\Gamma = 0.06 \gev \, ;
\\
m_t = 180 \gev \, , A=\mu = 250 \gev \, ,
&
\Rightarrow
&
\Gamma = 0.18 \gev \, ;
\\
m_t = 180 \gev \, , A=\mu = 1 \tev \, ,
&
\Rightarrow
&
\Gamma = 0.37 \gev \, .
\\
\end{array}
$$

The comparison between the present one-loop calculations and previous
approximations can be done by looking at fig.~2. One can see that the
`improved tree-level' result, obtained by using the tree-level formula
with one-loop-corrected values $\mh$, $\mhh$ and $\alpha$, can be off
by as much as a factor of $\sim 4$. The pure tree-level calculation would
be in general in much worse agreement. On the other hand, the effective
potential result is typically much closer to the full one. In particular,
the agreement betweeen the two methods is good for $\tb \gg 1$
or $m_H < 2 m_t, 2 \mstb$. As expected, radiative correction effects
are maximal for $\tb \sim 1$, corresponding to maximal top Yukawa
coupling for a given top mass. Also, momentum-dependent effects begin
to play a role only when the top or the stop thresholds are approached.
The effect of the top threshold is always smooth, but is nevertheless
clearly visible in figs.~2a--2d. Besides their effect on the
proper vertex, the stop thresholds give rise to
singularities in $\hat{\Pi}'_{HH}(m_H^2)$, corresponding to a breakdown of
perturbation theory near threshold: an example is the cusp appearing
in fig.~2d, whose details should therefore not be trusted.

To allow a better understanding of the origin of our results, we describe
in more detail the sources of numerically large corrections.
Wave-function renormalization effects in eq.~(\ref{aform})
are in general negligible, also thanks to the use of the one-loop-corrected
mixing angle $\alpha$ in the definition of the $(H,h)$ basis\footnote{
For example, had we used the tree-level definition of $\alpha$, the
$\Gamma_{Hh}(p^2)$ appearing in eq. (\ref{aform}) would have an expression
different from (\ref{offd}) and would contain large corrections,
proportional to $\mt^4$: this would jeopardize the validity of the
perturbative expansion underlying eq.~(\ref{aform}).}.
Large corrections come only from the proper
vertex $\Lambda_{Hhh}$, with the imaginary part never very large but
sometimes non-negligible, and diagrams of type (II) usually give small
contributions, unless $H$ is close to a stop threshold.

We conclude with some comments on the phenomenological implications of our
results. To consistently examine the effects of these corrections on the $H$
branching ratios, one should also compute the remaining partial widths at the
one-loop level, which goes beyond the aim of the present paper. At present,
only approximate computations exist, but some qualitative considerations are
nevertheless possible. For example, one might wonder if the phenomenological
analyses of $H$ signals at hadron colliders [\ref{kz}--\ref{pheno}], which
used the effective potential approximation in the computation of $\gh$, are
going to be significantly affected. The answer is negative: $H \to hh$ is
an important decay mode only for small $\tb$ and $m_H \simlt 2 m_t,
2 \mstb$, and in this region the effective potential approach is a rather
accurate approximation. The residual small corrections are negligible
compared with the uncertainties in the evaluation of the production
cross-sections. However, one could think that, in the happy event
that  $H$ is discovered, the effects studied in the present paper might
be important for a detailed study of its properties at a high-energy
and high-luminosity $e^+ e^-$ collider.

\section*{Acknowledgements}
We thank G.~Altarelli, R.~Barbieri, W.~Beenakker, F.~Caravaglios,
W.~Hollik, Z.~Kunszt, P.~Langacker and G.~Ridolfi for useful
discussions and suggestions.
\newpage
\appendix
\section*{Appendix}
We give here the explicit expressions of the coefficients
$c_{i\sqa\sqb}$ and $c_{ij\sqa\sqb}$:
\bean
c_{H\tilde{t}_1\tilde{t}_1} & = &
- \left[ d^t_{11} \mz^2 \cos(\alpha+\beta)
+ \frac{\mt^2\sin\alpha}{\sin\beta} + 2 s_t c_t B_{tH} \right]
\, ,
\\
c_{H\tilde{t}_2\tilde{t}_2} & = &
- \left[ d^t_{22} \mz^2 \cos(\alpha+\beta)
+ \frac{\mt^2\sin\alpha}{\sin\beta} - 2 s_t c_t B_{tH} \right]
\, ,
\\
c_{H\tilde{t}_1\tilde{t}_2} & = &
- \left[ d^t_{12} \mz^2 \cos(\alpha+\beta)
+ (c_t^2 - s_t^2) B_{tH} \right]
\, ,
\\
c_{H\tilde{b}_1\tilde{b}_1} & = &
- \left[ d^b_{11} \mz^2 \cos(\alpha+\beta)
+ \frac{\mb^2\cos\alpha}{\cos\beta} + 2 s_b c_b B_{bH} \right]
\, ,
\\
c_{H\tilde{b}_2\tilde{b}_2} & = &
- \left[ d^b_{22} \mz^2 \cos(\alpha+\beta)
+ \frac{\mb^2\cos\alpha}{\cos\beta} - 2 s_b c_b B_{bH} \right]
\, ,
\\
c_{H\tilde{b}_1\tilde{b}_2} & = &
- \left[ d^b_{12} \mz^2 \cos(\alpha+\beta)
+ (c_b^2 - s_b^2) B_{bH} \right]
\, ,
\\
c_{h\tilde{t}_1\tilde{t}_1} & = &
\left[ d^t_{11} \mz^2 \sin(\alpha+\beta)
- \frac{\mt^2\cos\alpha}{\sin\beta} + 2 s_t c_t B_{th} \right]
\, ,
\\
c_{h\tilde{t}_2\tilde{t}_2} & = &
\left[ d^t_{22} \mz^2 \sin(\alpha+\beta)
- \frac{\mt^2\cos\alpha}{\sin\beta} - 2 s_t c_t B_{th} \right]
\, ,
\\
c_{h\tilde{t}_1\tilde{t}_2} & = &
\left[ d^t_{12} \mz^2 \sin(\alpha+\beta)
+ (c_t^2 - s_t^2) B_{th} \right]
\, ,
\\
c_{h\tilde{b}_1\tilde{b}_1} & = &
\left[ d^b_{11} \mz^2 \sin(\alpha+\beta)
+ \frac{\mb^2\sin\alpha}{\cos\beta} - 2 s_b c_b B_{bh} \right]
\, ,
\\
c_{h\tilde{b}_2\tilde{b}_2} & = &
\left[ d^b_{22} \mz^2 \sin(\alpha+\beta)
+ \frac{\mb^2\sin\alpha}{\cos\beta} + 2 s_b c_b B_{bh} \right]
\, ,
\\
c_{h\tilde{b}_1\tilde{b}_2} & = &
\left[ d^b_{12} \mz^2 \sin(\alpha+\beta)
- (c_b^2 - s_b^2) B_{bh} \right]
\, ;
\eean
\bean
c_{Hh\tilde{t}_a\tilde{t}_b} =
\frac{1}{4}\sin 2\alpha
\left( 2 d^t_{ab} \mz^2 - \delta_{ab} \frac{m_t^2}{\sin^2\beta} \right)
\, ,
&  &
c_{Hh\tilde{b}_a\tilde{b}_b} =
\frac{1}{4}\sin 2\alpha
\left( 2 d^b_{ab} \mz^2 + \delta_{ab} \frac{m_b^2}{\cos^2\beta} \right)
\, ,
\\
c_{hh\tilde{t}_a\tilde{t}_b} =
\frac{1}{2} \left( d^t_{ab} \mz^2 \cos 2\alpha
- \delta_{ab} \frac{m_t^2 \cos^2 \alpha}{\sin^2\beta} \right)
\, ,
& &
c_{hh\tilde{b}_a\tilde{b}_b} =
\frac{1}{2} \left( d^b_{ab} \mz^2 \cos 2\alpha
- \delta_{ab} \frac{m_b^2 \sin^2 \alpha}{\cos^2\beta} \right)
\, .
\eean

The conventions for the squark masses and mixing angles, and the
symbols $d^q_{ab}$, $B_{qH}$, $B_{qh}$, etc., were all defined
in [\ref{b1},\ref{b2}]. The symbol $\delta_{ab}$ is the Kronecker delta.
Notice that the $c_{Hh\sqa\sqb}$ coefficients
disagree with those reported in ref.~[\ref{hunter}], as  already
observed in ref.~[\ref{pp}].

The function $F^c$, corresponding to the two-point scalar loop integral,
is given by
$$
F^c(m_1,m_2,m_3) = F(m_1,m_2,m_3)-i\pi\theta(m_3-m_1-m_2)
\sqrt{\left[1-\left(\frac{m_1+m_2}{m_3}\right)^2\right]
\left[1-\left(\frac{m_1-m_2}{m_3}\right)^2\right]}
\, .
$$

The function $f(m_a,m_b,m_c;p_1,p_2,p_3)$ corresponds to the three-point
scalar loop integral
\bea
f(m_a,m_b,m_c;p_1,p_2,p_3) & = &
\int_{0}^{1} d\!x \int_{0}^{x} d\!y
\left[ p_2^2 x^2 + p_1^2 y^2 + (p_3^2-p_1^2-p_2^2)xy \right.
\nonumber
\\
& +  &
\left. (-p_2^2+m_b^2-m_c^2)x + (p_2^2-p_3^2+m_a^2-m_b^2)y
+m_c^2 -i\epsilon \right]^{-1} \, .
\nonumber
\eea
It was studied and computed in ref. [\ref{thv}]. For an on-shell decay
its explicit expression can be written in the form
$$
f(m_a,m_b,m_c;p_1,p_2,p_3)  =
\frac{1}{D} \left[ Sp(A^+_{1+})+Sp(A^+_{1-})-Sp(A^-_{1+})-Sp(A^-_{1-})
+ (1 \to 2) + (1 \to 3) \right] \,  ,
$$
where $Sp(x)$ is the Spence function and
\bean
A^{\pm}_{1\pm} & \equiv &
\frac{\pm p_1^2-m_a^2+m_b^2+B_1}{B_1\pm C_1}
\, ,
\\
B_1  & \equiv &
\frac{1}{D} [ p_1^2(p_1^2-p_2^2-p_3^2+2m_c^2-m_a^2-m_b^2)
+(p_3^2-p_2^2)(m_a^2-m_b^2) ]
\, ,
\\
C_1 & \equiv &
\sqrt{p_1^4-2(m_a^2+m_b^2)p_1^2+(m_a^2-m_b^2)^2+i\epsilon}
\, ,
\\
D & \equiv &
\sqrt{p_1^4+p_2^4+p_3^4-2p_1^2 p_2^2-2p_1^2 p_3^2-2p_2^2 p_3^2}
\, .
\eean
The replacements $ (1 \to 2)$ and $ (1 \to 3)$ mean the following:
\bean
1 \to 2 & \equiv & (p_1, p_2, p_3, m_a, m_b, m_c)
\to (p_2, p_3, p_1, m_b, m_c, m_a)
\\
1 \to 3 & \equiv & (p_1, p_2, p_3, m_a, m_b, m_c)
\to (p_3, p_1, p_2, m_c, m_a, m_b)
\eean
Analogously to the function $F^c$, the function $f$ develops an imaginary
part above thresholds.
\newpage
\section*{References}
\begin{enumerate}
\item
\label{pioneer}
Y. Okada, M. Yamaguchi and T. Yanagida,
Prog. Theor. Phys. Lett. 85 (1991) 1;
\\
J. Ellis, G. Ridolfi and F. Zwirner, Phys. Lett. B257 (1991) 83;
\\
H.E. Haber and R. Hempfling, Phys. Rev. Lett. 66 (1991) 1815.
\item
\label{previous}
S.P.~Li and M.~Sher, Phys. Lett. B140 (1984) 339;
\\
J.F.~Gunion and A.~Turski, Phys. Rev. D39 (1989) 2701,
D40 (1989) 2325 and 2333;
\\
M.~Berger, Phys. Rev. D41 (1990) 225.
\item
\label{rga}
R. Barbieri, M. Frigeni and M. Caravaglios, Phys. Lett. B258 (1991) 167;
\\
Y. Okada, M. Yamaguchi and T. Yanagida, Phys. Lett. B262 (1991) 54;
\\
J.R. Espinosa and M. Quir\'os, Phys. Lett. B266 (1991) 389;
\\
M.A. Diaz and H.E. Haber, Phys. Rev. D45 (1992) 4246;
\\
K.~Sasaki, M.~Carena and C.E.M.~Wagner, Nucl. Phys. B381 (1992) 66.
\\
P.H. Chankowski, S. Pokorski and J. Rosiek, Phys. Lett. B281 (1992) 100.
\item
\label{epa}
R. Barbieri and M. Frigeni, Phys. Lett. B258 (1991) 395;
\\
J. Ellis, G. Ridolfi and F. Zwirner, Phys. Lett. B262 (1991) 477.
\item
\label{berz}
A. Brignole, J. Ellis, G. Ridolfi and F. Zwirner,
Phys. Lett. B271 (1991) 123 and (E) B273 (1991) 550.
\item
\label{drees}
M. Drees and N.M. Nojiri, Phys. Rev. D45 (1992) 2482;
\\
D.M.~Pierce, A.~Papadopoulos and S.B.~Johnson, Phys. Rev. Lett.
68 (1992) 3678;
\\
S. Kelley, J.L. Lopez, D.V. Nanopoulos, H. Pois and K. Yuan,
Phys. Lett. B285 (1992) 61.
\item
\label{da}
A. Yamada, Phys. Lett. B263 (1991) 233;
\\
P.H. Chankowski, S. Pokorski and J. Rosiek, Phys. Lett. B274 (1992) 191.
\item
\label{b1}
A. Brignole, Phys. Lett. B277 (1992) 313.
\item
\label{b2}
A. Brignole, Phys. Lett. B281 (1992) 284.
\item
\label{kz}
Z. Kunszt and F. Zwirner, preprint CERN-TH.6150/91, ETH-TH/91-7,
to appear in Nucl. Phys. B.
\item
\label{bbps}
V. Barger, M.S. Berger, A.L. Stange and R.J.N. Phillips,
Phys. Rev. D45 (1992) 4128.
\item
\label{pheno}
P.~Janot, Orsay preprints LAL-91-61 and LAL-92-27;
\\
H.~Baer, M.~Bisset, C.~Kao and X.~Tata, Phys. Rev. D46 (1992) 1067;
\\
J.F.~Gunion and L.~Orr, Phys. Rev. D46 (1992) 2052;
\\
J.F.~Gunion, R.~Bork, H.E.~Haber and A.~Seiden, Phys. Rev. D46 (1992) 2040;
\\
J.F.~Gunion, H.E.~Haber and C.~Kao, Phys. Rev. D46 (1992) 2907;
\\
H.~Baer, C.~Kao and X.~Tata, Florida preprint FSU-HEP-920717;
\\
H.~Baer, M.~Bisset, D.~Dicus, C.~Kao and X.~Tata, Florida preprint
FSU-HEP-920724;
\\
V. Barger, K. Cheung, R.J.N. Phillips and A.L.~Stange, Madison
preprint MAD-PH-696.
\item
\label{exper}
G.~Wormser, talk given at the XXVI Int. Conf. on High Energy
Physics, Dallas, Texas, USA, 1992, to appear in the Proceedings,
and references therein.
\item
\label{polish}
P.H.~Chankowski, S.~Pokorski and J.~Rosiek, Phys. Lett. B286 (1992) 307.
\item
\label{pp}
D.~Pierce and A.~Papadopoulos, Berkeley preprint UCB-PTH-92/23 and LBL-32498.
\item
\label{b3}
A. Brignole, unpublished, as quoted in ref.~[\ref{kz}].
\item
\label{hhn}
H.E.~Haber, R.~Hempfling and Y.~Nir, Phys. Rev. D46 (1992) 3015;
\\
H.E.~Haber and R.~Hempfling, Santa Cruz preprint SCIPP-91/33.
\item
\label{drbar}
W.~Siegel, Phys. Lett. B84 (1979) 193.;
\\
D.M.~Capper, D.R.T.~Jones and P.~Van~Nieuwenhuizen, Nucl. Phys. B167 (1980)
479;
\\
I.~Antoniadis, C.~Kounnas and K.~Tamvakis, Phys. Lett. B119 (1982) 377.
\item
\label{hunter}
J.F.~Gunion, H.E.~Haber, G.L.~Kane and S.~Dawson, {\em The Higgs Hunter's
Guide} (Addison-Wesley, New York, 1990).
\item
\label{thv}
G.~'t Hooft and M.~Veltman, Nucl. Phys. B153 (1979) 365.
\end{enumerate}
\vfill{
\section*{Figure captions}
\begin{itemize}
\item[Fig.1:]
The three basic topologies of the diagrams involving top, bottom, stop and
sbottom
\newline
exchanges and contributing to the one-loop $Hhh$ proper vertex ($q
= t,b$; $a,b,c, = 1,2$).
\item[Fig.2:]
The decay width $\gh$, as a function of $\mhh$, corresponding to the four
indicated parameter choices. Solid lines correspond to the full diagrammatic
calculation, dashed lines to the effective potential approach, dash-dotted
lines to the `improved tree-level' result.
\item[Fig.3:]
Contours in the $(m_A, \tan \beta)$ plane, corresponding to constant values
of $\Gamma (H \to hh)$, for the representative parameter choice $m_t = 140
\gev$, $m_{sq} = 1 \tev$, $A = \mu = 0$. The thick solid lines correspond
to $\Gamma (H \to hh)=0$ and in particular delimit the region where the
decay $H \to hh$ is kinematically disallowed.
\end{itemize}
}
\end{document}